\documentclass[aps,pra,twocolumn,amsmath,amssymb,floatfix,longbibliography]{revtex4-2}
	
\usepackage{amsmath}
\usepackage{txfonts}
\usepackage{microtype}
\usepackage{graphicx}
\usepackage{color}
\usepackage{ulem}
\usepackage{bm}
\usepackage[english]{babel}
\usepackage{braket}
\usepackage{mathtools,nicefrac}
\usepackage[caption=false]{subfig}
\DeclarePairedDelimiter{\abs}{\lvert}{\rvert}

\begin{document}

\title{The role of reflections in the generation of a time delay in strong field ionization}

\author{Daniel Bakucz Can\'{a}rio}
\email[]{daniel.bakucz-canario@mpi-hd.mpg.de}
\author{Michael Klaiber}
\author{Karen Z. Hatsagortsyan}
\email{k.hatsagortsyan@mpi-hd.mpg.de}
\author{Christoph H. Keitel}

\affiliation{Max-Planck-Institut f{\"u}r Kernphysik, Saupfercheckweg 1, 69117 Heidelberg, Germany}

\date{\today}

\def\be{\begin{equation}}
\def\ee{\end{equation}}
\def\bea{\begin{eqnarray}}
\def\eea{\end{eqnarray}}

\newcommand*\diff{\mathop{}\!\mathrm{d}}
\newcommand*\mystrut[1]{\vrule width0pt height0pt depth#1\relax}
\newcommand*\sech{\mathop{}\!\textnormal{sech}}
\newcommand*\vol{\Ket{\Psi^V_\mathbf{p}(t)}}
\newcommand*\bvol{\Bra{\Psi^V_\mathbf{p}}}
\newcommand*\intr{\textnormal{int}}
\newcommand*\sgn{\mathop{}\!\mathrm{sgn}}

\newcommand*\Ai{\mathop{}\!\mathrm{Ai}}
\newcommand*\Bi{\mathop{}\!\mathrm{Bi}}
\newcommand*\Erf{\mathop{}\!\mathrm{Erf}}

\newcommand*\mO{\mathop{}\!\mathcal{O}}

\begin{abstract}
The problem of time delay in tunneling ionization is revisited.
The origin of time delay at the tunnel exit is analysed, underlining the two faces of the concept of the tunnelling time delay: the time delay around the tunnel exit and the
asymptotic time delay at a detector. We show that the former time delay, in the sense of a delay in the peak of the wavefunction, exists as a matter of principle and arises due to the sub-barrier interference of the reflected and transmitted components of the tunneling electronic wavepacket. We exemplify this by describing the tunnelling ionization of an electron bound by a short-range potential within the strong field approximation in a ``deep tunnelling'' regime. If sub-barrier reflections are extracted from this wavefunction, then the time delay of the peak is shown to vanish. Thus, we assert that the disturbance of the tunnelling wavepacket by the reflection from the surface of the barrier causes a time delay in the neighbourhood of the tunnel exit.

\end{abstract}

\maketitle

\section{Introduction}

{\color{black}In both classical and quantum mechanics, the passage of time is always defined with respect to some dynamical variable (for instance, the hands of a clock). In the absence of a canonical quantum mechanical time operator, the definition and measurement of time in quantum mechanics is challenging since dynamical observables are inherently non-deterministic.
Despite the conceptual difficulties involved, operational techniques with which to measure time have been developed (see e.g. \cite{MacColl_1932,Peres_1980,Landauer_1994,Hauge_1998,Muga,deCarvalho,Davies_2005,Winful_2006,Sokolovski_2013} for overviews).

One of the main applications of these time measurement protocols is the study of the time taken for a particle to tunnel through a barrier, known as tunnelling time. Several definitions of the tunneling time exist, such as the Wigner time \cite{Eisenbud, EisenbudWigner, Smith_1960,wigner_1955,Czirjak_2000}, the B\"uttiker-Landauer time\cite{Butttiker_1982}, the Pollak-Miller time \cite{Pollak_1984}, the Larmor time \cite{Baz_1966,Rybachenko_1967,Buttiker_1983,Deutsch_1996,Ramos_2020}, the dwell time \cite{Nussenzveig_2000}, etc. \cite{Sokolovski_1987,Steinberg_1995,Yamada_2004,Landsman_2015,Maccone_2020}. Each definition corresponds to a specific aspect of the measurement process and, in general, these do not coincide.

The tunneling phenomenon plays an essential role in the strong field ionization process of atoms and in the related field of attoscience.} State-of-the-art techniques in attosecond science are now able to provide exceptional time and space resolution, reaching tens of attoseconds time- and Angstr\"om space-resolution. In particular, the attoclock technique provides time resolution of tunneling ionization\ \cite{Eckle_2008a}, and has inspired researchers to experimentally address the challenging problem of tunneling time \cite{Eckle_2008b, Pfeiffer_2012, Landsman_2014o, Sainadh_2019,Camus_2017}, i.e., how much time, if any, elapses during the quantum tunnelling process.

This question strikes at the fundamentals of quantum mechanics but experiments have often been followed by controversial discussion \cite{Eckle_2008b, Pfeiffer_2012, Landsman_2014o, Sainadh_2019, Camus_2017, Teeny_2016a, Teeny_2016b, Yakaboylu_2013, Han_2019, Orlando_2014, Orlando_2014PRA, Lein_2011,
 Landsman_2014, Torlina_2015, Ni_2018b, Ni_2018a, Ni_2016, Landsman_2015,Zheltikov_2016,Zimmermann_2016, Liu_2017, Song_2017,Yuan_2017, Klaiber_2018, Eicke_2018, Douguet_2018, Bray_2018, Crowe_2018, Ren_2018, Tan_2018, Sokolovski_2018, Quan_2019, Douguet_2019, Hofmann_2019, Serov_2019, Wang_2019, Yuan_2019, Kheifets_2020, Kolesik_2020} in the strong field community arguing either for, or against, the existence of a tunnelling induced time delay between ionised electron wavepacket and  ionising laser field.

In this respect {\color{black} we emphasise the distinction between two concepts of the tunneling time in strong field ionization}, namely, the time delay near the tunnel exit (the classically expected coordinate for the tunneled electron to emerge), and the asymptotic time delay {\color{black}\cite{Han_2019}}. While the latter is relevant to attoclock experiments, the former, known also as the Wigner time delay, can be calculated theoretically and measured in a Gedanken experiment with a so-called virtual detector \cite{Feuerstein_2003, Wang_2013}. In both cases, time delay arises due to sub-barrier dynamics.

The asymptotic time delay is derived from the asymptotic photoelectron momentum distribution (PMD). It is defined by the  classical backpropagation of the peak of the photoelectron asymptotic wavefunction up to the tunnel exit \cite{Torlina_2015,Ni_2016,Ni_2018b,Ni_2018a}.  A theoretical description of the experimental asymptotic time delay is challenging because of the entanglement of Coulomb field effects with those of the tunneling delay. {\color{black} In a regime far from over-the-barrier ionization (OTBI), the asymptotic time delay is vanishing \cite{Yakaboylu_2013, Han_2019} However, in regimes approaching the OTBI the asymptotic delay is not negligible; it was in fact found to be negative \cite{Torlina_2015,Ni_2018b,Ni_2018a}} and shown to arise due to interference of direct and the under-the-barrier recolliding trajectories \cite{Klaiber_2018}.

In the tunnelling region, the tunneling time delay is deduced by following the peak of the electron wavefunction, the so-called Wigner trajectory. A few de-Broglie wavelengths away from the tunnel exit, the electron dynamics are quasiclassical and the classical trajectory is accurately described by the eikonal of the wave function in the Wentzel-Kramers-Brillouin (WKB) approximation. However, near the tunnel exit, the WKB approximation fails and {\color{black} a full quantum mechanical description is given by the Wigner trajectory. Near the tunnel exit this description diverges strongly from the classical one but, nevertheless approaches it asymptotically, as one moves toward a detector.}
The Wigner time delay has been calculated numerically with the virtual detector method in Refs.~\cite{Yakaboylu_2013,Teeny_2016a,Teeny_2016b,Camus_2017}, showing positive time delay and nonvanishing longitudinal velocity at the tunnel exit.

In this paper we investigate the Wigner time delay in tunnelling ionization, which a virtual detector in a Gedanken experiment would observe. We concern ourselves chiefly with understanding the principles of the tunnelling delay, and to this end we consider a simple, time-dependent model of a one-dimensional (1D) atom, with an electron bound by a short-range potential, which is ionized by a half-cycle laser pulse.

At first sight, such abstraction may seem overly simplistic; nonetheless, it takes into account all the necessary features of tunnelling ionization. Firstly, ionization occurs mainly in the direction of the electric field so a 1D treatment is appropriate. Likewise, no real laser pulse is a half-cycle sinusoid but {\color{black}by considering one such laser we ensure no continuum electron-to-atomic core recollisions take place, which convolute the physical picture}. Such an interpretation is rendered clearer by having analytic expressions, unattainable with a pure Coulomb atomic potential but feasible with a short range potential.

In this manner, the time-dependent wavefunction is described within the strong field approximation (SFA).  The components of the wavefunction which are reflected by the tunnelling barrier are identified by analyzing the saddle points of the time-integral of the wavefunction. We show that the interference of reflected and transmitted components of the wavefunction generates the Wigner time delay, while the same delay vanishes when the reflected components are neglected. Additionally, the scalings with respect to the laser field strength of the Wigner time delay and group velocity at the tunnel exit are derived and interpreted.

We work well below the OTBI regime, in what we term the \textit{deep tunneling regime}. {\color{black}While in this regime the asymptotic tunneling time delay is vanishing  we show that the Wigner time at the tunnel exit (calculated in the first-order SFA) is not only nonzero, but sizeable, as already noted in }{\color{black}\cite{Han_2019}}. This is in contrast to the asymptotic time delay, where the first-order SFA result is vanishing and the second-order SFA is required to obtain a nonvanishing asymptotic time delay (attributable to sub-barrier recollisions and interference of paths).

The structure of the paper is as follows: in Sec.~\ref{section_ii} we introduce the basic ionization model,
calculate the SFA time-dependent wavefunction, and  construct the Wigner trajectory from the latter. To elucidate our method of analysing the contribution of reflections to the wavefunction, in Sec.~\ref{section_iii} we consider a simpler, analytic, model of ionization: an electron in a constant electric field.
The analyticity of this model allows us to relate the reflected components of the wavefunction to the contribution of the specific  saddle points of the integral representation of the wavefunction.
We apply this concept to the SFA wavefunction in Sec.~ \ref{section_iv}, whereby we show that neglecting reflections results in a zero Wigner time delay. Lastly, in Sec.~\ref{section_v}, we summarize {\color{black} and interpret} our results and discuss the implications of these on the interpretation of attosecond streaking experiments. {\color{black}For the reader unacquainted with strong field physics,} the Wigner time delay for scattering by a square potential barrier was previously analyzed in \cite{Bohm,Davies_2005}, where the significance of reflections to the time delay was alluded to. In Appendix \ref{appendix_a} this problem is revisited and the role of reflections is explicitly presented. {\color{black}Atomic units (a.u.) are used exclusively throughout this work.}

\section{Time Delay in Strong Field Ionization}
\label{section_ii}

The Wigner time, developed for scattering problems \cite{Eisenbud, EisenbudWigner, wigner_1955},  considers the time taken for the peak of the wavepacket to travel a given distance. For tunneling through a time-independent barrier, the Wigner time can be derived via the energy derivative of the phase of the tunnelling wavefunction~$\psi$:
\be
\label{wigdef}
\tau_W (x) = -i\,\frac{\partial}{\partial \,\mathcal{E}}\ln \frac{\psi (x)}{| \psi (x)| }.
\ee
{\color{black} The peak of the ionized wave packet leads to the asymptotic formation of a well-defined peak in the attoclock PMD, which is subsequently measured and interpreted.}

As the peak of the wavepacket follows the trajectory $\tau_W (x)$, the Wigner group delay velocity of the wave packet can be defined
\be
\label{wigvel}
v_W(x)=\left(\frac{\partial\tau_W (x)}{\partial x}\right)^{-1}.
\ee
When a  monochromatic wave is incident on a finite potential barrier, the propagation inside the barrier is a superposition of an exponentially suppressed wave (transmission) with a growing exponential wave, namely the reflection by the surface of the barrier (see e.g. Appendix \ref{appendix_a}). The goal of this paper is to establish the causal effect of wavefunction refections inside the barrier on the time delay in strong field ionization.

In order to do this, we calculate  the time-dependent ionized electron wavefunction at the position where the electron appears in the continuum, thereby allowing us to deduce the peak of the wavepacket with respect to time. {\color{black}The Wigner time delay can be calculated as the energy derivative of the phase of the wave function via Eq.~\ref{wigdef} in a static field but in a time-dependent laser field this prescription fails as energy is not conserved.} Correspondingly, we derive the Wigner time delay at the tunnel exit explicitly as the delay of the peak of the wavepacket, rather than from the energy derivative of the phase of the wavefunction.

\subsection{The Model}

Consider an electron initially bound in a 1D short-range potential
\be
V(x) = -\kappa \: \delta(x)
\ee
with a binding energy $I_p = \kappa^2/2$, and with a corresponding ground state wavefunction $\psi_0(x,t)= \sqrt{\kappa} \exp\left(-\kappa |x|+i I_p t \right)$.

To describe the generation of the peak of the ionized wavepacket at the tunnel exit, not the usual PMD at a detector, a half-cycle laser field is modelled, avoiding recollisions in the continuum. The laser electric field,
\begin{eqnarray}
\label{efield}
E(t)&=& -E_0 \cos^2(\omega t),
\end{eqnarray}
is switched on at $\omega t_0=-\pi/2$. {\color{black} In a half-cycle laser field the ionization happens mostly near the peak of the field. Here, nonadiabatic effects are governed by the second-derivative of the field at the peak $E''(0)/E(0)=2\omega^2\equiv \omega_{eff}^2$, where we define the effective frequency $\omega_{eff}=\sqrt{2}\omega$.}

We consider the tunnel ionization regime and keep  constant the Keldysh parameter \cite{Keldysh_1965}

\be
\gamma=\sqrt{\frac{I_p}{2 U_p}}\ll 1,
\ee

where {\color{black}$U_p=E_0^2/(4\,\omega_{eff}^2)$} is the ponderomotive potential.  
Thus, when varying the field strength $E_0$, we vary the frequency $\omega= \gamma E_0/ (\sqrt{2}\kappa)$ accordingly.

Moreover, we consider the so-called deep tunneling regime, wherein $E_0\ll E_{th}$, and  $E_{th}$ is the threshold field of OTBI. This threshold can be estimated as the field strength where the coordinate-saddle point of the SFA-matrix element, i.e. the starting point $x_s\approx \sqrt{\kappa/E_0}$ of the quantum orbit, becomes comparable the tunnel exit $x_e$, which for the short range potential corresponds to the condition $E_{th}\approx \kappa^3/4 $ \cite{Klaiber_2017A}.

The superposition of the laser field with the atomic potential creates a potential barrier, through which an electron can tunnel, with a characteristic classical tunnel exit $x_e = I_p/E_0 = 10$ a.u. {\color{black}\footnote{{\color{black} A more precise definition for the tunnel exit can be given by the SFA, namely  $x^{SFA}_e=\int_{t_s}^{\rm Re\{t_s\}}(A(t')-A(0))\;dt' \approx (1-\gamma^2/4)\, I_p/E_0= 0.995\, I_p/E_0$, where $t_s$ is the temporal saddle point of the SFA integral. The minuscule discrepancy between the classical tunnel exit, $x_e$, and its SFA correction for the parameters in this work ($\gamma=\sqrt{2}/10$) leads us to use the simpler of the two definitions.}}}. While this is a rudimentary estimate, it is in agreement with a tunnel exit derived quantum mechanically in our parameter regime {\color{black}as shown in Section~\ref{tunnelSFA}}.

\begin{figure*}
\subfloat[]{\includegraphics[scale=1.]{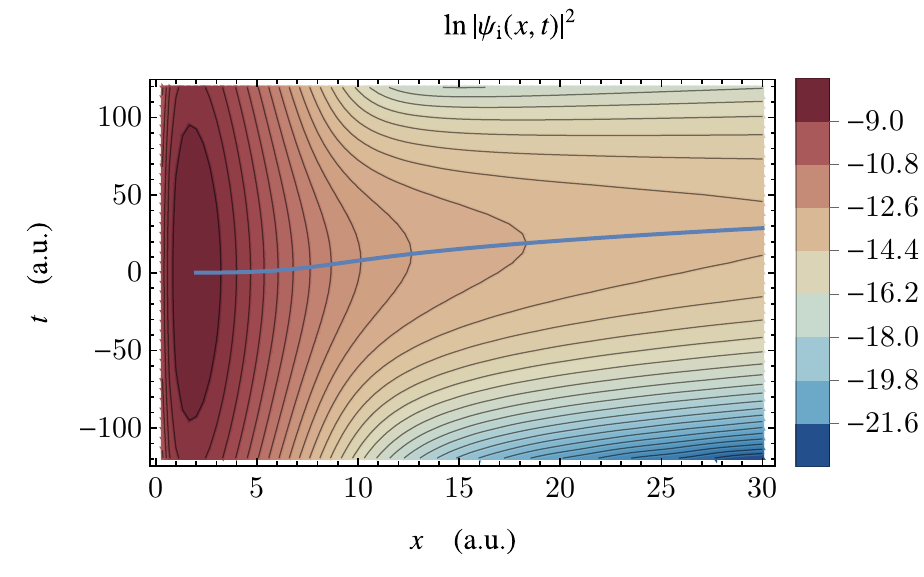}}
\subfloat[]{ \includegraphics[scale=1.]{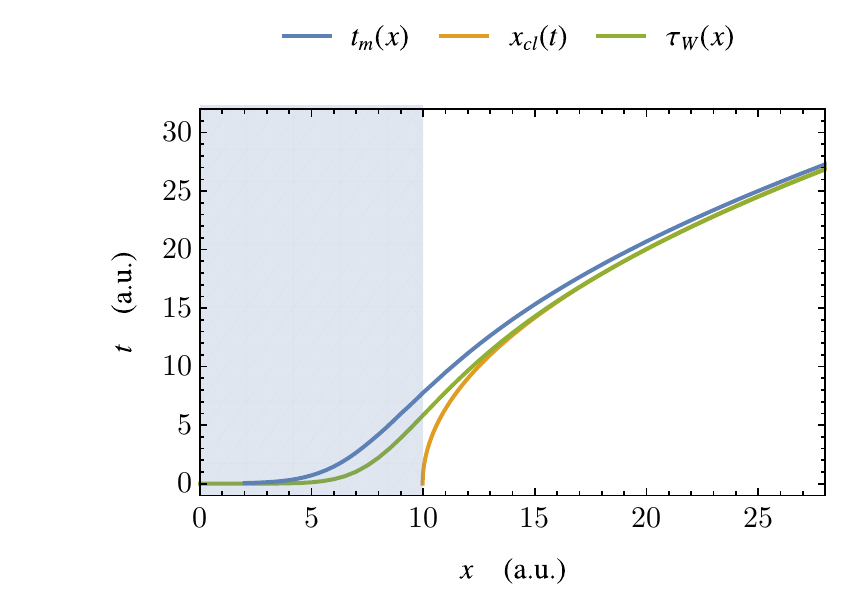}}
 \caption{\label{fullwf} \textbf{(a)} Ionization amplitude $|\psi_i(x,t)|^2$ as defined in Eq.~(\ref{int}). At every cross-section in $x$, the temporal peak of the wavefunction was determined (plotted in blue) which can be interpreted as the trajectory of the peak. \textbf{(b)} {\color{black} Comparison of trajectories in the region near the tunnel exit, $x_e$. The displayed curves are: the peak trajectory (blue, given by the maximum of $|\psi_i(x,t)|^2$), the classical trajectory (orange, given by the Newton equation starting at the tunnel exit), and the Wigner trajectory (green, given by Eq.~(\ref{WignerE})).} Under the barrier ($x<10$, shaded blue), there is an increasing delay of the peak w.r.t the laser peak. Outside the barrier, the peak trajectory rapidly converges with the classical and Wigner electron trajectory. {\color{black} Near the atomic core ($x<2$), the bound state, $|\psi_0(x)| \gg |\psi_i(x)|$, dominates the full wavefunction $\psi(x)=\psi_0(x)+\psi_i(x)$ so the meaning of a trajectory for ionization is lost.}}
 \end{figure*}
 \begin{figure}
\includegraphics[scale=1.]{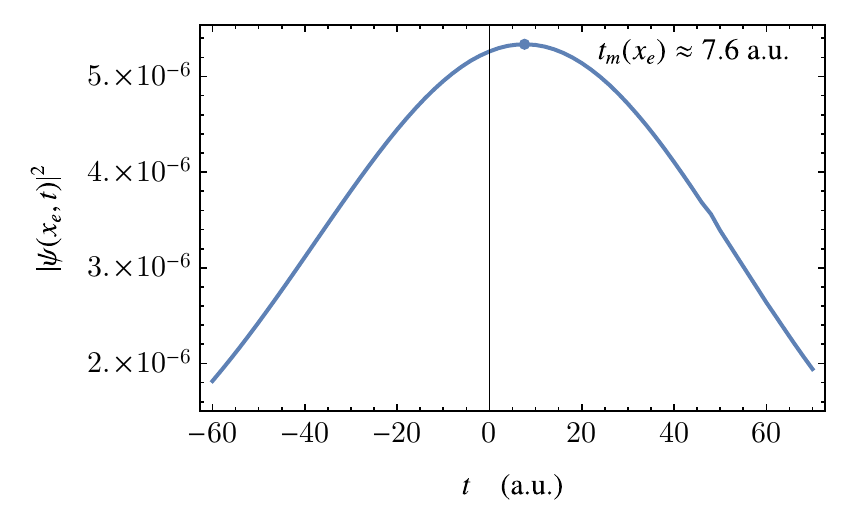}
 \label{distributions}
  \caption{Probability distribution $|\psi_i(x_e,t)|^2$ vs time  at the classical tunnel exit $x_e=I_p/E_0$. This distribution is peaked at a greater time, $t_{m}\approx 7.6$ a.u. $\approx 183$ as, marked by the dot, than the peak of the laser pulse. }
  \end{figure}

\subsection{Time Evolution of the Wavefunction}

Our investigation is based on the SFA wavefunction describing the electron dynamics in tunneling ionization. We define the Wigner trajectory and tunneling time delay employing the virtual detector approach \cite{Teeny_2016a,Teeny_2016b,Feuerstein_2003, Wang_2013} based on the SFA wavefunction.
The ionization dynamics are described by the Schr\"odinger equation
\be
i\frac{\partial}{\partial t}\Psi(x,t)= (H_0 +H_i)\Psi(x,t),
\ee
with the atomic Hamiltonian
\be
H_0 = -\frac 1 {2} \frac{\partial^2}{\partial x^2} + V(x),
\ee
and the interaction Hamiltonian with the laser field
\be
H_i =   x E(t)\;
\ee
The standard SFA is employed for the solution of the Schr\"odinger equation.
In the interaction picture with Hamiltonian $H=H_0+H_i$, the time evolution operator $U(t,t_0)$
satisfies the Dyson integral equations \cite{Becker_2002}
 \be
 U(t,t_0) = U_0(t,t_0) - i \int^t_{t_0} \: \diff t' \: U(t,t') H_i(t') U_0(t',t_0),
 \ee
where $U_0$ is the evolution operator corresponding to the Hamiltonian $H_0$. A perturbation series can be constructed by replacing the full evolution operator $U(t,t_0)$ in the Dyson integral with the evolution operator corresponding to the Hamiltonian $H_i$, namely
\be
U_f(t,t')= \int \diff p \ket{\Psi_p(t)}\bra{\Psi_p(t')},
\ee
where $\ket{\Psi_p}$ are the  Volkov states \cite{Volkov_1935}.  Each order in the perturbation theory corresponds to a sub-barrier recollision in our half-cycle laser field. As each sub-barrier recollision implies a longer tunnelling path, each term in this series is suppressed by the Keldysh exponent $\exp(\nicefrac{-2\kappa^3}{3E_0})$ and so, in contrast to the near OTBI regime, in the deep tunnelling regime a first order expansion in the SFA suffices {\color{black}(i.e. neglecting sub-barrier recollisions)} \cite{Klaiber_2018}.

Consequently, the electron state during the interaction takes the form $\ket{\psi(t)}=  \ket{ \psi_0(t)} + \ket{\psi_i(t)}$ where $\ket{ \psi_0(t)}$ is the eigenstate of the atomic Hamiltonian $H_0$ and
\begin{eqnarray}
\label{wavefunction}
\ket{\psi_i (t)}&=&-i \int^\infty_{-\infty} dp\int^t_{t_0} dt' \ket{\Psi_p(t)}\braket{\Psi_p(t')|H_{i}(t')|\psi_0(t')}, \qquad
\end{eqnarray}
{\color{black} describes the electron dynamics induced by the laser field, which  generates the Wigner trajectory. It is the dominant part of the wavefunction at distances far from the core, and therefore, we concentrate on the analysis of its behavior.}
\begin{figure*}
\begin{center}
\label{scalings}
\subfloat[\label{time}]{\includegraphics[scale=1.]{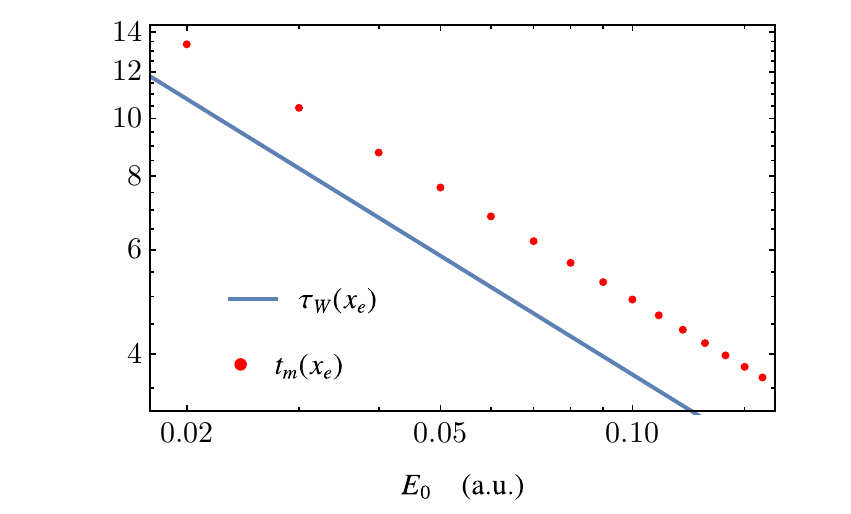}}\hfill
\subfloat[\label{momentum}]{\includegraphics[scale=1.]{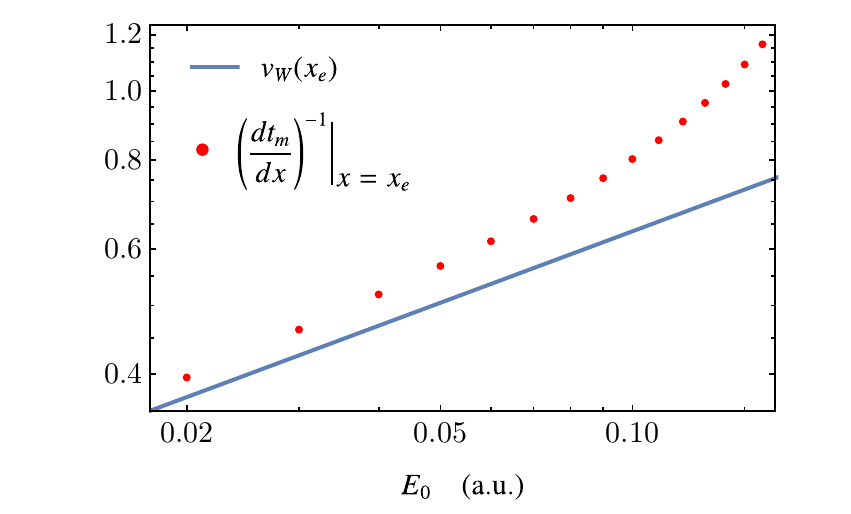}}
\caption{\label{tmaxE0} Log-log plots of the scaling w.r.t. field strength,  $E_0$, of the \textbf{(a)} Wigner time delay and \textbf{(b)} group velocity at the classical tunnel exit, $x_e=I_p/E_0$, for the time evolved SFA wavefunction (red dots) and the adiabatic constant field wavefunction (blue lines). For the time evolved wavefunction, the maxima in time, $t_m$, of the probability distribution $|\psi(x_e,t)|^2$ and its derivative were calculated; for the adiabatic case, results follow from Eqs.~(\ref{tauxe}) and (\ref{vxe}). The agreement in the trends is good, deteriorating as one approaches the OTBI threashold $E_{th}\approx 0.25$ a.u. }
\end{center}
\end{figure*}
Denoting the electronic kinetic momentum as
\be
{\cal P}(t)\equiv p+ A(t),
\ee
with the laser vector potential $A(t)=-\int_{-t_0}^t E(t') \,dt'$, the 1D Volkov state in the length gauge is
\be
\Ket{\Psi_p (t)} =  \Ket{{\cal P}(t)}e^{-i \, S(t)}
\ee
with a plane wave component $\Braket{x|p} =(2\pi)^{-\frac12} \exp({i\, p \,x}) $ and the  Volkov-phase
\be
S(t)= \frac{1}{2 }\int^t \diff \tau \, {\cal P}(\tau)^2\;.
\ee
We can expand the overlap in (\ref{wavefunction}) using a resolution of identity in a dummy variable $x'$
\be
\braket{\Psi_p(t')|H_{i}(t')|\psi_0(t')}= \frac{e^{i\left(I_p t'-S(t')\right)}}{\sqrt{2\pi}} E(t')\int^\infty_{-\infty} dx'\, e^{i\,\mathcal{P}(t')x'-\kappa |x'|} x'
\ee
While this integral is soluble, we do not evaluate it but rather first integrate (\ref{wavefunction}) w.r.t. to $p$ and then $x'$. With the notation $\Delta f_n(t,t') = f_n(t)-f_n(t')$, and defining the integrals $f_n(t)=\int^t dt' A(t')^n $, we may perform the initial integral over $p$ which is a simple Gaussian integral. The resulting expression

\begin{widetext}
\be
\psi_i(x,t)=   \int^t_{t_0} \diff t' \int_{-\infty}^\infty \diff x' \frac{x' \sqrt{\kappa}\,E(t')}{\sqrt{2\pi\,i (t-t')}}\exp\left[i\left(I_p t' +i\kappa |x'|+x(A(t)-A(t'))+\frac {i\left((x-x')-\Delta f_1(t,t')\right)^2}{2(t-t')} -\frac {\Delta f_2 (t,t')}2)\right)\right]
\ee

may then be finally integrated in $x'$ to yield

\begin{align}
\label{int}
\psi_i(x,t)&=  \int^t_{t_0} d t'\,\sqrt{\kappa} \, E(t') \left(\frac{z_-^2 (1+\Erf(z_-)) e^{-z_+^2} }{z-\kappa
   }+\frac{z_+^2(1- \Erf(z_+)) e^{-z_-^2}}{z+\kappa}\right) \exp \left(i \zeta \right)
 \equiv \int^t_{t_0} d t' \exp({ i\, \Phi(x,t,t')})
 \end{align}
\end{widetext}
where we have defined $z_\pm = \sqrt{\frac{i}{2}(t-t')}  (z\pm\kappa)$ for
\be
z= -i \, \left(A(t' )+ \frac{x-\Delta f_1 (t,t')}{t-t' } \right)
\ee
and where
\begin{align}
\zeta&=I_p t'+ x\,A(t)+\\
   &(t-t' ) \left(\frac{(x-\Delta f_1(t,t'))^2}{2 (t-t' )^2}+\kappa ^2+z^2\right)-\frac{\Delta f_2(t,t')}{2}
\end{align}

Our aim is to calculate the  wavefunction and Wigner time delay in the interaction region around the tunnel exit by studying the phase $\Phi$. Thus, in deviation to standard SFA studies, the time integral is calculated up to the finite observation time $t$, meaning the integration must be performed numerically.  The usual saddle-point integration method is valid only for sufficiently large values of $t$, placing one well outside the tunnelling region.

\subsection{Numerical Integration}

For the numerical integration of the SFA wavefunction Eq.~(\ref{int}) an electric field strength $E_0=0.05$ a.u. and {\color{black}Keldysh parameter $\gamma=\sqrt{2}/10\approx0.14$} were chosen to ensure deep tunnelling is considered. {\color{black}With these parameters, our half-cycle laser pulse spans $[-314, 314]$ a.u.} We present results for the case of hydrogen, for which $I_p=\kappa^2/2=1/2$ a. u., implying an OTBI threshold field $E_{th}\approx 0.25$ a.u. The time integration was carried out with the standard numerical integration routine of \textit{Mathematica} 12 to a precision of 30 digits.

The space-time probability distribution, $|\psi_i(x,t)|^2$, is presented in Fig.~\ref{fullwf}. From this distribution  the Wigner trajectory is derived as follows: the probability time-distribution at each space point $x$ is invoked (see Fig.~\ref{distributions} for the case of the tunnel exit coordinate $x=x_e$) and the maximum $t_{m}(x)$ of this distribution is derived. The Wigner trajectory is represented by the function $t_{m}(x)$, which runs along all maximum points of the space-time distribution. {\color{black} Near the core, $|x|\lesssim 3$ a.u., the bound state dominates the wavefunction $|\psi_0| \gg |\psi_i|$, so the concept of the Wigner trajectory loses its meaning.

We underline that the wavefunction $\psi_i$ vanishes after the laser field is switched off. Thus, $\psi_i$ does not include the possible net excitations of the atomic states due to interaction with the laser field. This is a feature of the SFA, where the exact time-evolution operator in the final state is replaced by the Volkov propagator, representing the continuum electron in the laser field.}

We note the maximum of the wavefunction displays a time delay with respect to the peak of the field. The tunneling time delay under the barrier ($x<10$) increases when moving towards the tunnel exit. Outside the barrier, this trajectory rapidly approaches the classical electron trajectory (beginning at the classical tunnel exit with zero momentum). The slight deviation is due to quantum mechanical corrections to the quasiclassical wave function not far from the tunnel exit.

The main idea advocated in this paper is that the Wigner time delay during tunneling is closely related to reflections arising during tunneling dynamics. It is straightforward in the simple case of tunneling through a box potential to show that reflections are responsible for the tunneling time delay, as is done in Appendix \ref{appendix_a}. However, unlike in the separable box potential case,  a given wavefunction is not easily deconstructible as a simple superposition of reflected and transmitted components corresponding to simple decaying /growing exponentials. Instead, the transmitted and reflected components are completely encapsulated in the wavefunction making it rather more problematic to disentangle. This is ultimately achieved in Sec.~\ref{section_v} for the SFA wavefunction presented in this paper but in order to highlight the main aspects of the method we discuss beforehand a simpler example, that of tunneling in a constant electric field.

\section{Time delay in a constant field}
\label{section_iii}

The adiabatic model of ionization, namely atomic ionization in a constant field, will be helpful for our purposes as it is analytically tractable. Moreover, it will provide us with a reference with which to compare our earlier time dependent model.

\subsection{Ionisation in an adiabatic field}

Consider a bound electron of energy $-I_p$ in 1D $\delta$-potential ionized by a constant electric field $E_0$. The continuum eigenstate of the electron in this field is given by the  solution to the time-independent Schr\"{o}dinger equation:
 \be
\label{eq:schrodinger}
-\frac{1}{2} \frac{\diff^2 \psi}{\diff x^2} +\left(I_p  -E_0\:x  \right) \psi(x)=0,
\ee
which has as a general analytical solution as a superposition of the Airy functions of the first and second kind
\be
\psi(x)= c_1 \Ai(\tilde{x}) + c_2 \Bi(\tilde{x}),
\ee
where
\be
\tilde{x}=\left(\frac{2}{E_0^2}\right)^\frac{1}{3}\left(I_p-E_0\:x\right).
\ee
The requirement that the  wavefunction be a travelling wave as $x\rightarrow \infty$ imposes the condition $c_1 = -i c_2$
so that the wavefunction takes the form
\be
\label{constwf}
\psi(x)=T\left[\Bi(\tilde{x})+i\,\Ai(\tilde{x})\right],
\ee
where $T$ can be determined by matching this wavefunction to the bound-state solution of the atom;  this prefactor plays a role in the ionization amplitudes but is irrelevant to the phase of the wavefunction. We  calculate the Wigner  time delay for this wavefunction via the energy derivative of the phase
\begin{align}
\tau_W(x) = i \frac{\partial}{\partial {I_p}} \ln \frac{\psi}{|\psi|}
= \frac{2^{\frac13}}{\pi  E_0^{\frac23}} \;\frac{1}{
   \Ai(\tilde{x})^2+\text{Bi}(\tilde{x})^2}.
   \label{WignerE}
\end{align}
The constant field model allows one to estimate the scaling of the Wigner time delay, {\color{black} given by Eq.~(\ref{wigdef})}, at the tunnel exit $x_e=I_p/E_0$:
\be
\label{tauxe}
\tau_W(x_e)=\frac{3^{4/3} \, \Gamma \left(\frac{2}{3}\right)^2}{2^{5/3} \pi} \frac{1}{E_0^{2/3}}.
\ee
{\color{black}Using Eq.~(\ref{wigvel}),} we derive the scaling of the Wigner group velocity of the electron  at the tunnel exit,
\be
\label{vxe}
v_W(x_e)   = \frac{2 \left(\frac{2}{3}\right)^{2/3} \sqrt{\pi }\, \Gamma
   \left(\frac{7}{6}\right)}{\Gamma \left(\frac{2}{3}\right)^2} \,E_0^{1/3},
\ee
and find it consistent with the estimate of the scaling of the electron momentum $p_e \sim E_0 \, \tau_W$ in Ref.~\cite{Klaiber_2018}.

We  compare  the Wigner time delay and the group delay velocity in the constant field case with the exact time dependent SFA calculations in Fig.~\ref{tmaxE0}. {\color{black} The time delay in the SFA scales with the electric field in a similar manner to the adiabatic case, $\tau_W(x_e)\propto 1/E_0^{2/3}$, but shifted in time by a constant.} This shift stems from the fact that the Wigner trajectory is derived via the {\color{black} energy derivative of the phase of the wave function (which corresponds to the following the peak of the wave packet, neglecting its spreading)}, while the time-integration responsible for the formation of the ionization wave packet in SFA is calculated exactly numerically.

{\color{black} The scaling of the Wigner group velocity using the SFA is also in agreement with the  adiabatic estimate, $v_W(x_e)\propto E_0^{1/3}$. However, the relative error between these two grows with the field strength, becoming appreciable at higher field strengths when approaching the OTBI threshold. This is consistent with the simple estimate for the threshold for OTBI, $E_{th} \approx 0.25$ a.u., where a tunnelling description no longer applies.}

Note that the Wigner trajectory in the constant field case is determined from the solution of the time-independent Schr\"odinger equation (energy eigenstate), while  in the SFA case from the time-dependent wavefunction.
In the first case, the Wigner trajectory is defined as the derivative of the wavefunction phase with respect to the energy, which corresponds to determining the coordinate of the peak of the electron wavepacket at a fixed time moment, i.e. determining the motion of the wavepacket peak. In contrast, in the SFA case we explicitly determine the maximum of the wavefunction  in time for a fixed coordinate.

\begin{figure}
\begin{center}
\includegraphics[scale=1.]{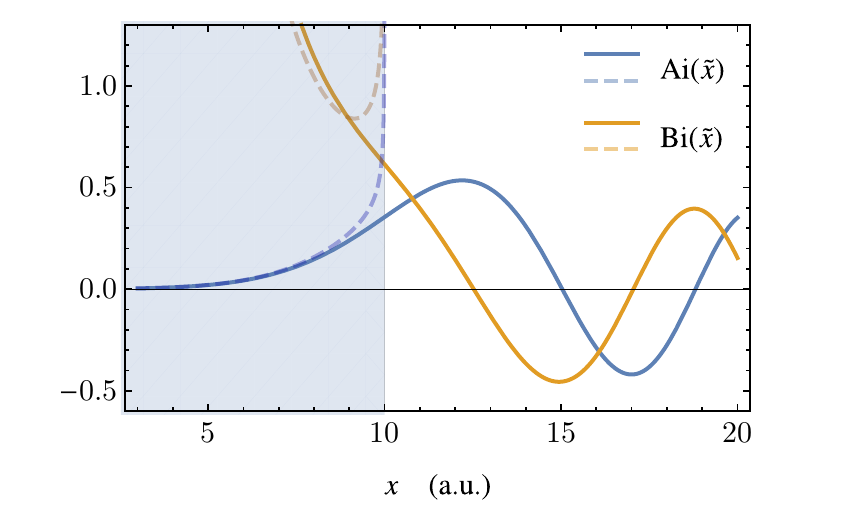}
\end{center}
\caption{ The solution to the electron in a constant field problem is given by a superposition of Airy functions of the first and second kind, $\Ai(\tilde{x})$ and $\Bi(\tilde{x})$, plotted in blue and orange respectively. These functions have very accurate asympotic expansions, shown dashed, which can be derived by considering the saddle points of the Airy integrals (\ref{Ai})-(\ref{Bi}). Under the barrier, $x<10$ (shaded blue), these expansions show that the wavefunction components $\Ai(\tilde{x})$ and $\Bi(\tilde{x})$ respectively correspond to the reflected and transmitted components of the wavefunction.}

\label{airyfunctions}
\end{figure}

\begin{figure*}
\begin{center}
\subfloat[\label{airycontours}]{
\includegraphics[scale=1.]{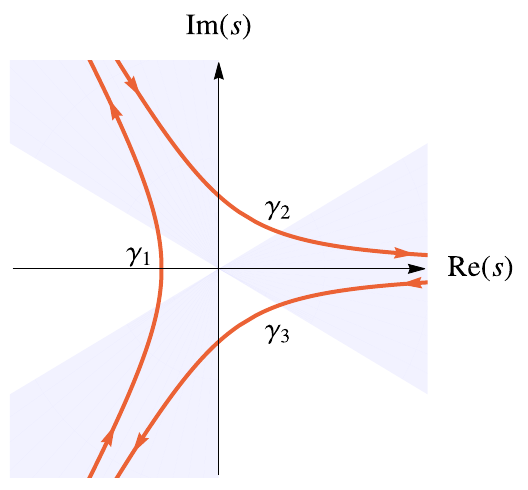}}
\hfill
\subfloat[\label{subfig:x<0}]{
\includegraphics[scale=1.]{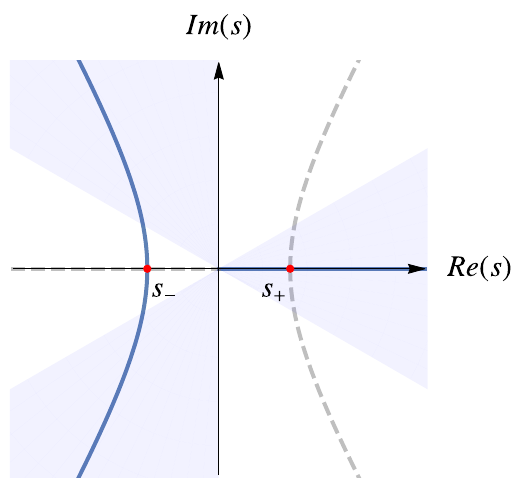}}
\hfill
\subfloat[\label{subfig:x>0}]{
\includegraphics[scale=1.]{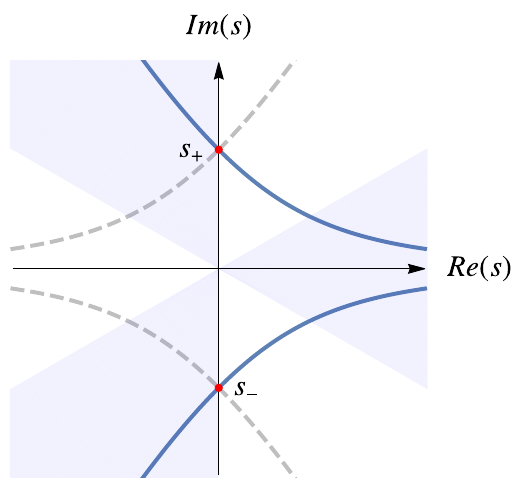}}
\caption{ The Airy integral $\int_\gamma \, ds \exp(\tilde{x}\,s-s^3/3)$  is defined on the complex $s$ plane; it converges when the endpoints of its infinite contour, $\gamma$, lie in the shaded areas. The canonical contours defining the Airy function of the first ($\gamma_1$) and second ($\gamma_2-\gamma_3$) kind are shown in \textbf{(a)}.  Contributions to the Airy integral arise principally from portions of the contour near the saddle-points of the integrand function $\exp(\tilde{x}\,s-s^3/3)$. The configuration of the saddles $s_\pm(x) = \pm \sqrt{\tilde{x}}$ (red dots) in the complex plane is shown for positions \textbf{(b)} inside the barrier{ \color{black}($\tilde{x}>0$)}, and \textbf{(c)} outside the barrier {\color{black}($\tilde{x}<0$)}.
{\color{black}Through each saddle point pass two perpendicular level curves, the path of steepest descents (solid blue) and the path of steepest ascents (dashed gray)}. Deforming the defining contours in \textbf{(a)} through the steepest descents contours in \textbf{(b)} and \textbf{(c)} provides asymptotic expansions for the Airy function, as in Eqs.~(\ref{aiasym}) and (\ref{biasym}).}
\label{saddleairy}
\end{center}
\end{figure*}

\subsection{Reflections in constant field}

To investigate the effect of under-the-barrier reflected components, we need to isolate their contribution to the full wavefunction. The analytical wave function of Eq.~(\ref{constwf}) consists of a superposition of Airy functions $\Ai(\tilde{x})$ and $\Bi(\tilde{x})$ which can be given interpretation as the under-the-barrier reflected and transmitted components of the tunneling wavefunction, respectively.

This interpretation can be generally understood by considering the Airy functions, as shown in Fig.~\ref{airyfunctions}. The $\Bi$-component of the wavefunction decays exponentially as it approaches the tunnel exit, $x=x_e$, from the atomic core at $x=0$, and hence can be seen to correspond to the transmission component; likewise, the exponentially growing $\Ai$-component corresponds to wavefunction reflection.

This interpretation can be formally established by considering the integral representation of the Airy functions
\begin{align}
\label{Ai}
\Ai(\tilde{x})&= \frac1{2\pi i} \int_{\gamma_1} ds \; \exp(\tilde{x}\,s-\frac{s^3}3),\\[17pt]
\label{Bi}
\Bi(\tilde{x})&= \frac{1}{2\pi} \int_{\gamma_2 -\gamma_3} ds \; \exp(\tilde{x}\,s-\frac{s^3}3),
\end{align}
where the complex plane integration paths $\gamma_i$ are indicated in Fig.~\ref{saddleairy}~(a). These integrals converge when their endpoints lie in the slices of the complex plane defined by $-\frac \pi 6 < \theta < \frac \pi 6$, $\frac \pi 2 < \theta < \frac{5 \pi}6$, and  $\frac {7\pi}6 < \theta< \frac{3 \pi} 2 $, where in polar $\{r,\theta\}$ coordinates $s= r \exp(i \, \theta)$.
In these regions, shaded blue in Fig.~\ref{saddleairy}, the integrand vanishes rapidly as $r\rightarrow \infty$.

The majority of the contributions to the Airy integrals thus come from around the saddle points,
\begin{align}
\label{spm}
s_{\pm} = \pm \sqrt{\tilde{x}},
\end{align}
of the argument of integrand. The Airy contours, $\gamma_i$, can be deformed into paths that to go through these saddle-points and, using the standard technique of saddle-point integration method \cite{Bender}, asymptotic expressions for the Airy integrals can be determined.

The saddle-points, and the respective paths of steepest descents {\color{black} and ascents}, are illustrated in Figs.~\ref{saddleairy}~(b) and (c); since these are dependent on $\tilde{x}$ the application of the saddle point method is different for the two cases of inside {\color{black}($\tilde{x}>0$)} and outside {\color{black}($\tilde{x}<0$)} the potential barrier.
As shown in Fig. ~\ref{saddleairy}~(b), for {\color{black}$\tilde{x}>0$} we may deform the contour $\gamma_1$ smoothly into the path of steepest descents for the saddle point $s_{-}$ and in doing so obtain the asymptotic approximation
\begin{align}
\label{aiasym}
\Ai(\tilde{x})&=  \frac{\exp\left(-\frac23 \tilde{x}^{\frac32}\right) }{2 \,\sqrt{\pi}\, \tilde{x}^{\frac14}}+\mO(\tilde{x}^{-\frac32}).
\end{align}
Likewise, we may deform the contours $\gamma_2$ and $-\gamma_3$ to both go through the steepest descent path of the saddle point $s_+$ and hence deduce the asymptotic form of the $\Bi(x)$-function under the potential barrier
\begin{align}
\label{biasym}
\Bi(\tilde{x})&=  \frac{\exp\left(+\frac23 \tilde{x}^{\frac32}\right) }{\sqrt{\pi}\, \tilde{x}^{\frac14}}+\mO(\tilde{x}^{-\frac32}).
\end{align}
From a tunnelling particle's perspective, one may identify the $\Ai(x)$ function and its saddle-point $s_-$ with the reflected part of the wavefunction, and the saddle-point $s_+$, or the $\Bi(x)$ function with the transmitted part of the wavefunction: the transmitted wave decays as it moves to the tunnel exit, while the reflected wave propagates from the exit decaying exponentially toward the atomic core, as shown in Fig.~\ref{airyfunctions}.

{\color{black}In the following section, this relation between sub-barrier reflections and the contribution of one saddle point to the time-integrand of the wavefunction amplitude is applied to the problem of ionization in a time-dependent field. This then allows us to investigate the role of sub-barrier reflection in the formation of a tunneling time delay.}

\section{Wigner Time and Reflections in the SFA}
\label{section_iv}

 \begin{figure*}
   \includegraphics[scale=1.]{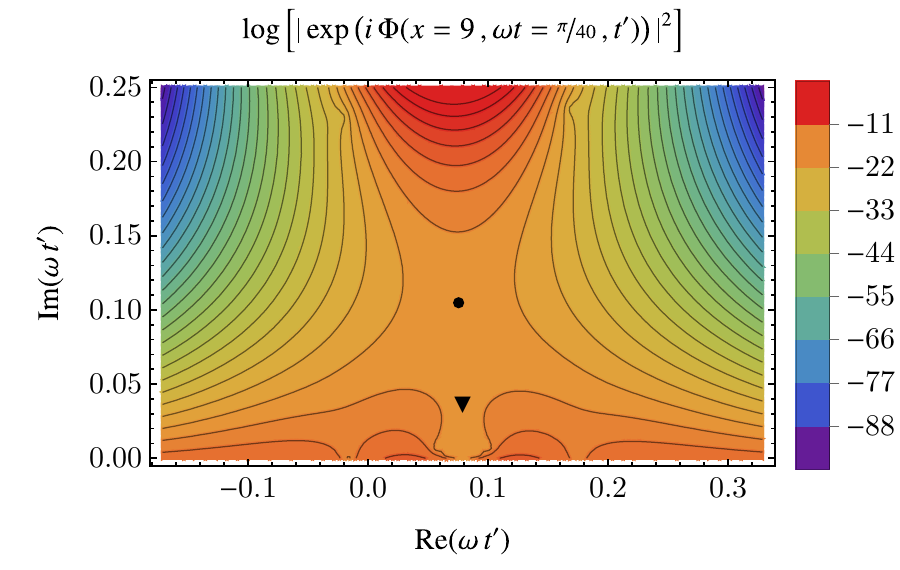}
 \caption{\label{steepest1} Complex $t'$ plane of the argument $\Phi(x,t,t')$ of the wavefunction integral  $\psi_i=\int^t_{t_0} dt' \exp(i\,\Phi)$, for parameters $x=9$ and $\omega t=\pi/40$. For these configurations there are two saddle-points, denoted by a circle and triangle, the former which, $t_+$, corresponds to reflections.  The saddle-point method can be used to solve the wavefunction integral, when the path is taken over both saddle-points. We use a partial path given by (\ref{wavefunctionnr}) to remove the contribution of reflections by integrating over only one of the saddle-points. Other possible configurations of the saddle points are shown in Fig~\ref{contourgrid}.}
 \end{figure*}

As shown in Sec.~\ref{section_iii}, we may reasonably ascribe the saddle-point of the integrand of the wavefunction to under the barrier reflection or transmission. To reveal the contribution of the reflection to the tunneling time delay, we investigate the complex continuation of the integrand function of the SFA wavefunction, Eq.~\ref{int}. The insight developed previously, namely the relation of a specific saddle-point of the integrand function to reflection-like behaviour in the wavefunction, is used to extract the contribution of reflections to the wavefunction by modifying the path of integration the complex plane.

\begin{figure}
   \includegraphics[scale=1.]{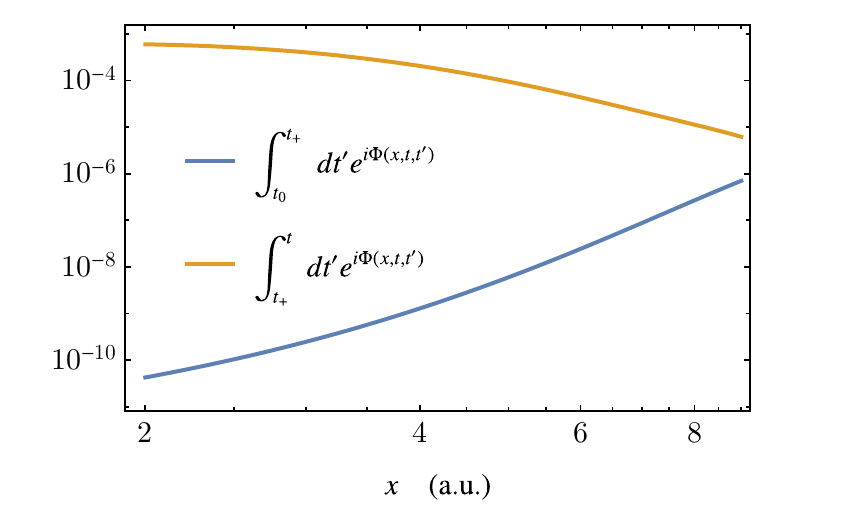}
 \caption{\label{exponentials} Log-log plot of the relative contributions to the wavefunction amplitude from the portion of the integration path up to the saddle point (blue) and from the saddle point to the given real time $t$ (orange) for various positions $x$ under the classical barrier (i.e. far from the atomic core, but smaller than the tunnel exit). The behaviours are approximately decaying and growing exponentials, respectively, analogous to transmitted and reflected parts of the wavefunction.}
\end{figure}

\begin{figure*}
\begin{center}
\subfloat[]{\includegraphics[scale=1.]{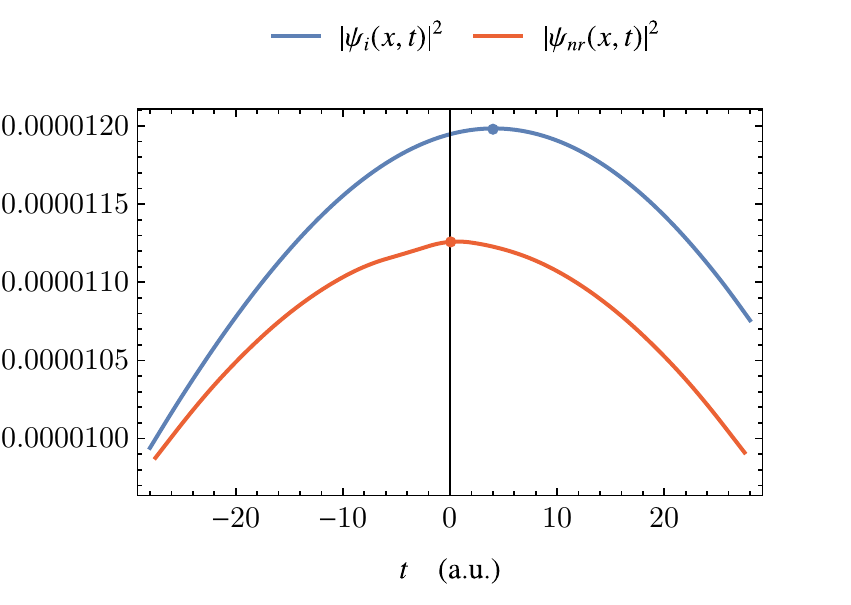}}\hfill
\subfloat[]{\includegraphics[scale=1.]{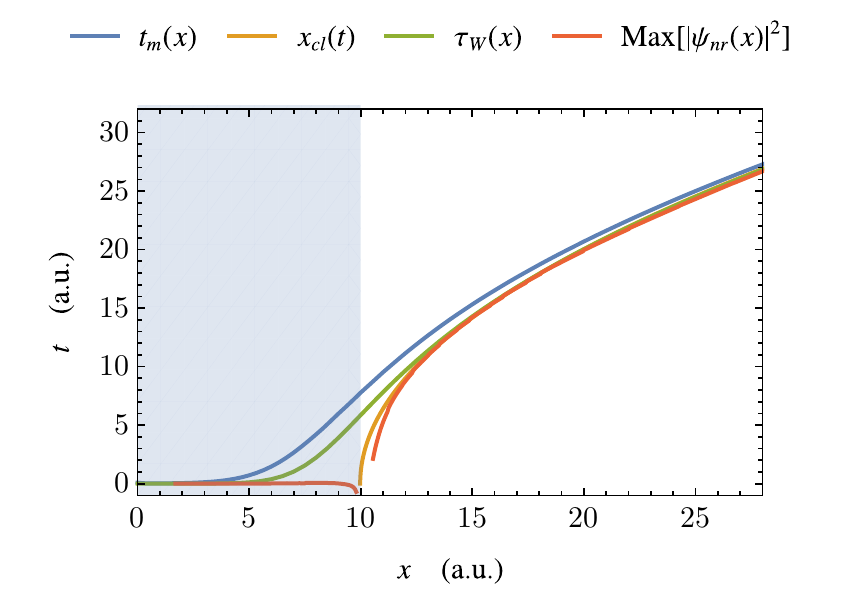}}
     \caption{\label{refl}\textbf{(a)} Temporal probability distributions at the position $x=8 \textnormal{ a. u.}$ (under the barrier, near tunnel exit), for the ionised wavefunction, $\psi_i$, and the pseudo-wavefunction neglecting reflections, $\psi_{nr}$. The physical wavefunction has accrued a Wigner time delay with respect to the laser field, whereas the maximum of the pseudo-wavefunction is synchronous with the laser field peak. \textbf{(b)} Trajectories in the $(x,t)$ plane via: the maximum of the wavefunction $\psi_{i}$ (given by $t_m(x)$), the maximum of the wavefunction neglecting reflections $\psi_{nr}$, and the energy derivative of the constant field wavefunction $\tau_W(x)$. The classical electron trajectory, $x_{cl}(t)$, starting at the tunnel exit at the peak of the laser field is also shown. Under the barrier, the probability distribution neglecting reflections $|\psi_{nr}|^2$ shows zero time delay. In regions where Eq~(\ref{cond}) does not apply, mostly near the classical tunnel exit, we may not identify reflections nor plot a subsequent trajectory. }
 \end{center}
    \end{figure*}

In general, the complex $t'$-plane picture of the phase $\Phi$ in Eq.~(\ref{int}) has many similarities to that of adiabatic ionisaion. However, it depends continuously not only on the coordinate $x$ but now also on observation time $t$, making its analysis somewhat more involved. In adiabatic ionisation, there was only one degree of freedom, $\tilde{x}$, and the saddle points were either purely real or imaginary.

As shown in Fig.~\ref{steepest1}, in the time dependent case there are also two saddle-points of relevance in the SFA integral denoted $t_\pm$. Their arrangement in the complex $t'$-plane is dependent on $x$ and $t$ but, as in the adiabatic case, the real coordinate $x$ determines whether they are vertically or horizontally aligned (in or outside the barrier); the observation time $t$ merely shifts the axis of symmetry which is always observed around the line $t'=t$. A more detailed discussion of the configuration of saddle points in $(x,t)$ parameter space may be found in Appendix~\ref{appendix_contourgrid}.

\subsection{Extracting Reflections}

In the adiabatic case, a partial integration over only contour containing the saddle point $s_-$ would exclude the exponentially growing contributions to the wavefunction, i.e. the reflections, associated with the saddle point $s_+$.

We use the same principle in the time dependent case; in Fig.~\ref{steepest1} we identify the saddle point corresponding to reflections, denoted $t_+$ (represented by a circle), and the secondary saddle point $t_{-}$ (represented by a triangle). When we consider the wavefunction under the potential barrier, we can split the integration contour in Eq.~(\ref{int}) {\color{black} into two parts: from the start of the pulse, $t_0=-\pi/(2\omega)$}, to the upper saddle point, $t_+$, and from $t_+$ to the real time $t$.

The contributions of these two contours to the wavefunction are shown in Fig. \ref{exponentials}. The former integral has the form of a growing exponential and so can ostensibly be identified as the reflected part of the wavefunction while the latter is a decaying exponential identifiable as the transmitted part. To neglect contributions from the reflections to the full wavefunction we identify the following wavefunction:
\begin{align}
\label{wavefunctionnr}
\psi_{nr}(x,t)=\begin{dcases} \quad \int_{t_0}^{t_+} dt'\exp(i\, \Phi(x,t,t')) & x>x_e\\[14pt]
\quad \int^t_{t_+} dt' \exp(i\, \Phi(x,t,t')) & x<x_e\\
 \end{dcases}
\end{align}
Our method of distinguishing the reflection contribution is based on the distinguishability of the saddle-points contributions to the wavefunction integral. This is not possible when the two saddle points are so close that the cubic term in the expansion of the phase $\Phi(x,t,t')$ becomes non-negligible \cite{Chester_1957}, i.e. when
 \be
 \label{cond}\abs*{\frac{\frac{\partial^3}{\partial t'^3}\Phi(x,t,t')}{\left(\frac{\partial^2}{\partial t'^2}\Phi(x,t,t')\right)^{\frac32}}}_{t'=t_\pm}  \gtrsim 1.
 \ee
for either saddle point $t_\pm$. This occurs in the region very near the barrier boundary and at the peak of the laser pulse, as discussed further in Sec.~\ref{tunnelSFA}. In this case, the reflection cannot be separated in the wavefunction in a meaningful manner.

We find that when reflections are extracted from the wavefunction using Eq.~(\ref{wavefunctionnr}) the Wigner time delay under the barrier vanishes, as evident in both panels of Fig.~\ref{refl}. That is, when reflections are neglected, there is no time delay between the peak of the electron probability distribution and the peak of the laser field. Moreover, outside the barrier, the trajectory of the peak rapidly becomes classical.

Plotted also in Fig.~\ref{refl}(b) is the trajectory for an electron in a constant field, given by Eq.~(\ref{WignerE}). For parameters in the deep tunnelling regime, the SFA trajectory (given by the peak of the wavefunction), and the classical trajectory, the constant field trajectory display very similar behaviour as the pulse elapses (the trajectories do not completely coincide far away due to the time dependence of the field and the need to match a tunnel exit, respectively).

At a detector infinitely far away such as one in an  attoclock experiment, a signature of time delay would be absent for the deep tunneling conditions detailed in this work. Thus, it is of importance to distinguish between a time delay of the peak of the wavefunction near the tunnel exit and an asymptotic time delay at a detector. Since in attoclock experiments the asymptotic momentum distribution is measured, i.e., the asymptotic time delay, one expects to find a zero time delay in the deep tunnelling regime. However, near the tunnel exit, quantum mechanical considerations must be taken into account and a quantum mechanical treatment such as the one exposed here exhibits an explicit non-zero Wigner delay.

\subsection{Tunnel Exit from the SFA}
\label{tunnelSFA}

We may also use the saddle point analysis of Eq.~(\ref{int}) to extract relevant physical information about the electron dynamics. In particular, and in analogy to the constant field scenario, the topology of the saddle points through the complex $t'$ plane reveal the functional behaviour of the electronic wavepacket.

\begin{figure}[b]
 \includegraphics[scale=1.]{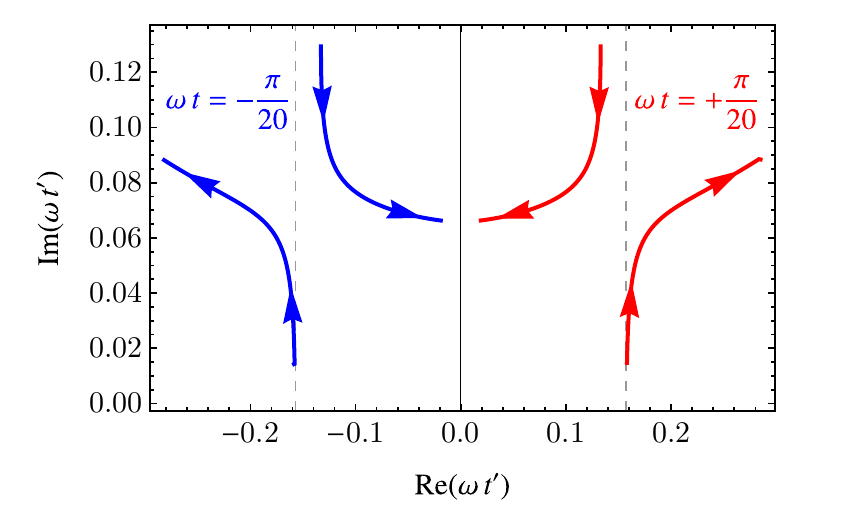}
 \caption{\label{sppaths1} Positions of the upper and lower saddle-points of the integral (\ref{int}) in the complex $t'$ plane with varying $x$ and for fixed $\omega\, t= \pm0.1 \pi/2$ (blue and red, respectively). The saddles draw smooth curves with varying $x$, where the arrows indicate growing values of $x$. For values of $x \lesssim 7$, the lower saddle-point disappears below the imaginary axis. Each pair of curves is centred around the line  $\textnormal{Re}(\omega t') = \omega t$, shown in dashed. For the case $\omega t=0$, not shown, the two lines meet one point corresponding to $x=x_t$.}
 \end{figure}

The paths of these two saddle point, for two representative choices of time $t$ while varying $x$, are shown in Fig.~\ref{sppaths1}, closely resembling those corresponding to Figs.~\ref{saddleairy} (b) and (c).
Much as in the constant field case, where the saddles are purely real or imaginary, as given by Eq.~(\ref{spm}) and Fig.~\ref{saddleairy}, the pair of saddle points in the SFA is aligned around the line given by $\textnormal{Re}[t']=t$; this alignment is either vertical or horizontal depending on the coordinate $x$. For a fixed observation time $t$, there corresponds a coordinate $x$ of closest approach between the saddle points. For the peak of the pulse, $t=0$ there exists a coordinate $x_t$ where the saddle points merge completely, as they do in the case $\tilde{x}=0$ for the constant field electron.

Thus, two lines, $t=0$ and $x=x_t$, define separatrices for the behaviour of the saddle points and their contributions to the integral: the behaviour of the wave function can be seen to change depending on whether $x>x_t$. This is a phenomenon exactly analogous to the merging of the saddle points for constant field case of Sec.~\ref{section_iii}: there, the merging of the saddle points occurs at the classical exit, $\tilde{x}=0$,  and separates the regions of evanescence and oscillation in the wave function.

We may use this parallel to identify the merging point as a measure of the tunnel exit. In our study, for a field strength of $E_0=0.05$ and {\color{black} $\gamma=0.14$}, we find that this exit takes the value $x_t \approx 10.2$ a.u.; this is in good agreement with the classically expected tunnel exit $x_e=I_p/E_0=10$ a.u. More precise agreement is achieved with the probability averaged tunnel exit
\be
x_K= \dfrac{\int dt' \frac{I_p}{|E(t')|} \exp\left(-\frac{2\kappa^3}{3|E(t')|}\right)}{\int dt' \exp\left(-\frac{2\kappa^3}{3|E(t')|}\right)}\approx 10.35 \;\textnormal{a.u.},
\ee
where the integration runs over the whole laser pulse.

\section{Conclusions}
\label{section_v}

In this work, we have analyzed the tunneling time delay in strong field ionization for a simple model of an atom with a short-range potential. The wave function was calculated to first-order in the SFA for any intermediate time, enabling a quantum treatment of dynamics in the region where quasiclassical descriptions break down, viz. around the classical tunnel exit. For a given coordinate, the peak of the wave packet shows a time delay with respect to the peak of the laser field and this time delay is positive at the classical tunnel exit.

We argue that reflections of the electron wavepacket under the tunneling barrier are fundamentally responsible for this non-zero time delay around the tunnel exit. This is a general phenomenon, present in any regime of strong field ionization, as well as in any tunneling process. In particular, a simple showcase is presented in Appendix \ref{appendix_a} for the case of tunneling through a box potential.

To identify and separate contributions of reflected components to the wavefunction, we first considered the analytically soluble case of an electron in a constant electric field. In addition to providing a benchmark to our SFA wavefunction, the constant field wavefunction can be deconstructed by analysing its definition as an integral in the complex plane; this integral is dominated by two saddle points, allowing us to unambiguously identify the transmitted and reflected wavefunction components as contributions to the integral from the regions around each saddle point.

We apply this observation to the SFA wavefunction by defining an integration contour that purposefully neglects the contribution of the reflection saddle point, and obtain a pseudo-wavefunction distribution that has zero time delay with respect to the peak of the laser field everywhere under the barrier. Moreover, the tunnel exit is unequivocally determined from the SFA theory as the spatial co-ordinate at which the saddle points merge at the peak of the laser field.

Finally, we emphasize the distinction between the time delay present at the tunnel exit (theoretically measurable using a virtual detector) and an (experimentally observed) asymptotic delay. We have shown within the first-order SFA that the signature of the time delay in the electron wavefunction vanishes since the ionized electron propagates to the detector as the classical (i.e. with zero time delay) and the quantum mechanical electron share equivalent asymptotic trajectories. Thus, the asymptotic time delay associated with the direct ionization path is always vanishing. Nevertheless, a more accurate estimate of the asymptotic time delay via the second-order SFA in Ref.~\cite{Klaiber_2018} has shown that interference of the direct and the sub-barrier ionization paths generates non-vanishing negative asymptotic time delay in the near-threshold of the OTBI regime, which fades out in the deep tunneling regime.

{\color{black}\section{Acknowledgements}

The authors are grateful to the referees for their useful remarks, which generated fruitful discussions. This article comprises parts of the PhD thesis work of Daniel Bakucz Can\'ario, submitted to Heidelberg University, Germany.}

\appendix

\section{Tunneling through a box potential barrier}
\label{appendix_a}

In this appendix we consider the role of quantum reflections in the tunneling time delay for an electron wavepacket tunnelling through a one dimensional box potential. This is perhaps the simplest example of tunnelling time delay since the wavefunction is readily separable (and of analytic solution) in the regions inside and outside the barrier, allowing the contributions of reflections in to the time delay to be clearly identified.

It should be mentioned that this model differs from true strong field ionization in that it considers the scattering of a wavepacket incident on a potential barrier; in actual ionization the electron originates \textit{from within} a potential barrier. Be that as it may, this study provides a simple, intuitive picture of tunnelling time delay, suitable even for the uninitiated.

\subsection{The Box Potential}
 \begin{figure}[h]
 \includegraphics[scale=1.]{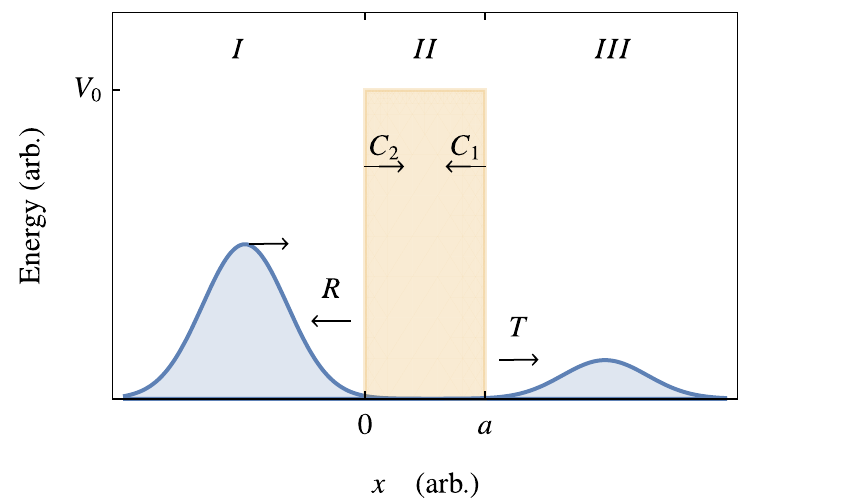}
 \caption{\label{diagram} Pictorial representation of the square barrier potential for an incident monochromatic wave. The wavefunction is a piecewise solution of the Schr\"{o}dinger equation for the three regions shown, given by Equations (\ref{wf})-(\ref{wf2}).}
 \label{box}
 \end{figure}
Consider a wave packet
\be
\Psi(x,t) = \int \diff p\: f(p-p_0)\: \psi(p) \: e^{-i \:E(p)\;t}
\ee
with energy $E(p)={p}^2/2$, incident on a potential barrier $V(x) = V_0$ for $0 \le x \le a$ and $0$ elsewhere, where $f(p-p_0)$ is some distribution peaked at $p_0$ (e.g. a Gaussian), as shown Fig.~\ref{box}. Each $p$-component wavefunction obeys the time-independent Schr\"{o}dinger equation with the piecewise solution
\bea
\label{wf}
\psi_{I}(x) &=& e^{i p \,x} + R e^{-i p\,x}, \\
\label{wf2}\psi_{II}(x) &=& C_1 e^{q x} + C_2 e^{- q x}, \\
\psi_{III}(x) &=& T e^{i p x},
\eea
with momenta $p = \sqrt{2\,E}$, $q = \sqrt{2(V_0-E)}$. The amplitude of the incoming wave has been set to unity, without loss of generality, and the co-efficients $C_1$ and $C_2$ are the typical reflection and transmission coefficients under the barrier, respectively.

Matching the above solutions and their derivatives at the boundaries yields the coefficients
\begin{align}
\label{B}
C_1&=\dfrac{(-2i\chi)(1+i\chi)e^{-\xi}}{(1-i \chi)^2 \:e^{\xi} - (1+ i \chi)^2\: e^{-\xi}},\\[16pt]
\label{C}
C_2&=\dfrac{(-2i\chi)(1-i\chi)e^{\xi}}{(1-i \chi)^2 \:e^{\xi} - (1+ i \chi)^2\: e^{-\xi}},  \\[16pt]
\label{R}
R &= \frac{(1 + \chi^2)\;\left(e^{-\xi}- e^{\xi}\right)}{(1-i \chi)^2 \:e^{\xi} - (1+ i \chi)^2\: e^{-\xi}}, \\[16pt]
\intertext{and,}
\label{T}
T&=\frac{(-4i\chi)e^{-i p a}}{(1-i \chi)^2 \:e^{\xi} - (1+ i \chi)^2\: e^{-\xi}}.
\end{align}
We have introduced the dimensionless parameters $\chi = p/q$ and $\xi = q\,a $ which, loosely speaking, determine the relative length and height of the barrier respectively.

\subsection{Time Delay }

The wavepacket after transmission is of the form
\be
\label{trans}
\Psi_{III} = \int \:\diff p\: | T(p) |\:\exp{[i \left( \varphi(p) + p x - E(p) t\right)]},
\ee
where $T=|T| e^{i \varphi}$. The maximum of this amplitude occurs when the phase in Eq.~(\ref{trans}) vanishes, that is when (after some re-arrangement):
\be
\label{xdelay}
x= p_0\, t - \left[\frac{\partial \varphi}{\partial p}\right]_{p=p_0}
\ee
In the absence of a potential barrier, the peak travels at the classical velocity (in atomic units) $p_0$. Equation (\ref{xdelay}) shows that the barrier causes a delay of the peak in reaching a given position $x$, a delay which is given the by the energy derivative of the transmission phase $\varphi$. Thus,
\be
\label{tau}
\tau = \frac{1}{p_0} \left[\frac{\partial \varphi}{\partial p}\right]_{p=p_0} + \frac{a}{p_0},
\ee
 gives the time delay of the peak after crossing the barrier, i.e. the tunneling time. For the  box potential, one finds

\be
\label{tau2}
\tau = \frac{a}{p_0} \frac{\frac{1}{2 \xi}\left(\chi+\frac{1}{\chi}\right)^2\tanh(\xi) + \frac{(1-\chi^2)}{2} \sech^2(\xi)}{1+ \frac{1}{4} \left(\chi - \frac{1}{\chi} \right)^2 \tanh^2(\xi)}.
\ee

\subsection{Quantum Reflections and Time Delay}

{\color{black} What is the origin of this time delay?}  For longer barriers, $\xi \gg 1$, the tunnelling time $\tau$ becomes
\be
\label{delay1}
\lim_{\xi \gg 1}\tau\approx \frac{2\chi}{p_0^2},
\ee
vanishing when $\chi\ll 1$. Thus, the time delay is vanishing for $\xi\gg 1$ and $\chi\ll 1$. With this knowledge, we can analyze the exact wave function inside the barrier. {\color{black}The reflection coefficient $C_1$ in the given limiting conditions, $\xi\gg 1$ and $\chi\ll 1$,
 \be
 C_1\approx-2i \chi e^{-2\xi},
 \ee
 vanishes asymptotically. This suggests an intuitive link between the reflection of wave components and the retardation of the wave packet.}

{\color{black}In the box potential case, in contrast to the tunneling problem, there exists, however, a second reflection: the reflection from incidence on the first surface of the barrier, described by the coefficient $R$.} Classically, $R=1$, and there is no time delay at reflection. However, in the quantum case $R-1\neq 0$, inducing a time delay of the electron wave packet from entry into the barrier. In the limit $\xi\gg 1$, and $\chi\ll 1$, we have $R
\approx -(1+2i\chi)$, i.e.,
  \begin{equation}
   |1-R|\approx \chi \ll 1
 \end{equation}
Analyzing the exact wave function, we can conclude that the time delay vanishes when the reflection from the barrier surface is classical, and when the reflection inside the barrier is negligible.

One can deduce the time delay caused by reflection at the entry by applying the Wigner delay formula, Eq.~\ref{wigdef}, at $x=0$. For $\xi \gg 1$, $R \approx -(1+i\chi)/(1-i\chi)$ and
\begin{align}
\tau_{entry}&= \frac{-i}{p_0} \left.\,\frac{\partial}{\partial p}\ln\left(\frac{\psi_I(x)}{| \psi _I(x)| }\right)\right|_{x=0}\approx\quad \frac {\chi} {p_0^2}
=\quad \frac 1 2  \;\lim_{\xi \gg 1}\,\tau. \;
  \end{align}
Thus, one can conclude the time taken for the peak of the electron wavepacket to travel through the box potential is caused in equal parts by the reflections of the wavepacket on the barrier entry ($x=0$) and exit ($x=a$) surfaces.

\section{Configurations of saddle points} \label{appendix_contourgrid}
\label{appendix_b}

\begin{figure*}
\begin{center}
\includegraphics[scale=1]{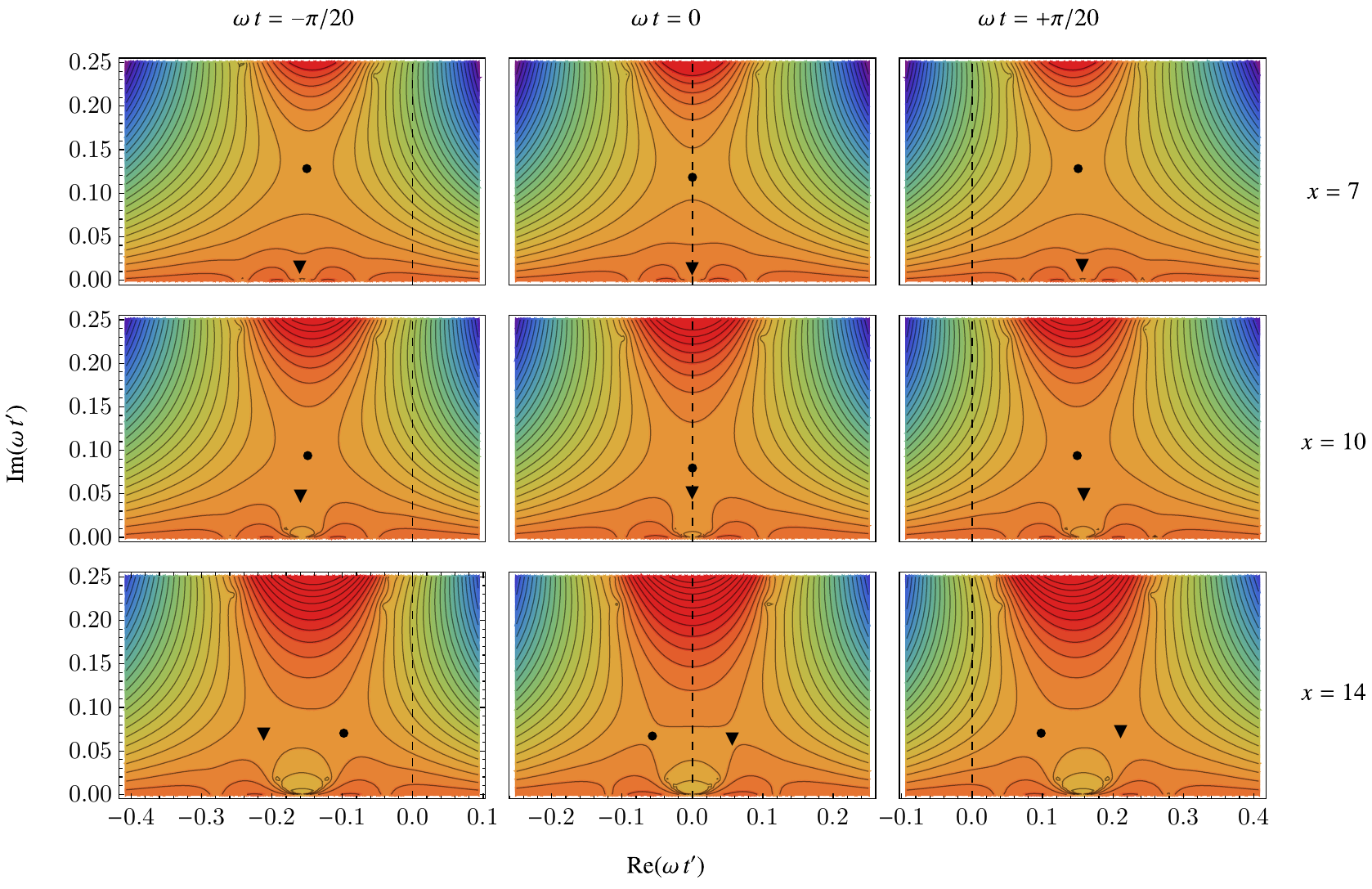}
\caption{Configurations of the saddle points of the argument $\Phi(x,t,t')$ of the wavefunction integral  $\psi_i=\int^t_{t_0} dt' \exp(i\,\Phi)$ in complex $t'$ plane, for parameter ranges $x=7, 10, 14$ and $\omega t=-\pi/20,0,+\pi,20$. The dashed line corresponds to $t'=0$ and each plot is centred around the (scaled) observation time $\omega t$.  The scale and colour coding are identical to those of Fig.~\ref{steepest1}. As $x$ increases the two saddle points approach vertically and, after a closest approach, separate horizontally. For the peak of the pulse, $\omega t =0$, this closest approach is zero and the two saddle points merge at the point $x_t$. Otherwise, varying $t$ skews the relative orientation of the saddle points around the line $\textnormal{Re}[t']=t$.}
\label{contourgrid}
\end{center}
\end{figure*}

The configurations of the saddle points of the argument $\Phi(x,t,t')$ of the wavefunction integral  $\psi_i=\int^t_{t_0} dt' \exp(i\,\Phi)$ in complex $t'$ plane, for a large $(x,t)$ parameter range are presented as a table in Fig.~\ref{contourgrid}. The laser pulse evolves from left to right in this figure and mostly shifts the reference line around which the saddle points are centred, namely $\textnormal{Re}[t']=t$. It should be noted, by the definition in Eq.~\ref{int}, a singularity is always to be observed at the point $t'=t$.

As one moves down the table, configurations of the complex $t'$ plane are shown for spatial coordinates inside, neighbouring, and outside the tunnelling barrier ($x=7, 10,$ and $14$ a.u. respectively). There are marked differences between each observation coordinate but the behaviour is reminiscent of the complex plane configuration for the Airy integral, displayed in Fig.~\ref{saddleairy}. Indeed, the two pictures can easily be reconciled by a $\nicefrac{\pi}{2}$ rad. clockwise rotation.

Inside the barrier, the saddle points are vertically aligned along the line $\textnormal{Re}[t']=t$. As one increases the observation coordinate and approaches the tunnel exit, $x\approx 10$, the two saddle points approach each other vertically; as one exits the region neighbouring the tunnel exit, $x \gg 10$, the saddles then separate from each other on the horizontal.

As one approaches the peak of the laser field, $t=0$, the distance of closest approach around the tunnel exit shrinks. \textit{At} the exact peak, this distance is zero; that is, the saddle points merge. This phenomenon is directly comparable to the merging of saddle points for the Airy integral at the classical tunnel exit. Thus, we are able identify a new tunnel exit, $x_t$, from the complex plane landscape, a topic discussed in Sec.~\ref{tunnelSFA}.

The identification the exact value for the co-ordinate $x_t$ to a desired precision is in principle achievable by a binary search or global minimization of the distance function for the saddle points. However, such precision was deemed unnecessary for the purposes of this work.

\bibliography{strong_fields_bibliography3}

 \end{document}